\documentclass[prl,showpacs,amssymb,floatfix,twocolumn]{revtex4-1}
\usepackage{amsmath}
\usepackage{graphicx}% Include figure files
\usepackage{epsfig}% Include figure files
\usepackage{dcolumn}% Align table columns on decimal point
\usepackage{bm}% bold math
\usepackage{clrscode}

\def\(({\left(}
\def\)){\right)}

\def\[[{\left[}
\def\]]{\right]}

\newcommand{\be}{\begin{equation}}
\newcommand{\ee}{\end{equation}}
\newcommand{\bea}{\begin{eqnarray}}
\newcommand{\eea}{\end{eqnarray}}

\begin{document}

\title{Belief-Propagation Guided Monte-Carlo Sampling}

\author {Aur\'elien Decelle$^{1}$ and Florent Krzakala$^{2,3}$,}
\affiliation{ $^1$ Dipartimento di Fisica, Universit\`a La Sapienza, a Piazzale Aldo Moro 5, I-00185 Roma, Italy. \\
  $^2$ CNRS and ESPCI ParisTech, 10 rue Vauquelin, UMR 7083 Gulliver, Paris 75000, France. \\
  $^3$ Laboratoire de Physique Statistique, CNRS UMR 8550, 
Université P. et M. Curie Paris 6 et École Normale Supérieure, 24, rue Lhomond, 75005 Paris, France.
}

\begin{abstract}
  A Monte-Carlo algorithm for discrete statistical models that
  combines the full power of the Belief Propagation algorithm with the
  advantages of a heat bath approach fulfilling the detailed-balance  is presented. First we extract randomly a
  sub-tree inside the interaction graph of the system. Second, given the boundary conditions, Belief
  Propagation is used as a perfect sampler to generate a
  configuration on the tree, and finally, the
  procedure is iterated. 
This approach is best adapted for locally tree-like graphs and we therefore tested it on random graphs for hard models such as spin-glasses, demonstrating its state-of-the art status in those cases.
\end{abstract} % AD a bit too much ??
  
% 64.70.qd Theory and modeling of the glass transition 89.70.Eg
% Interdisciplinary: Computational complexity 75.50.Lk Magnetic
% properties of materials: Spin glasses and other random magnets

\maketitle

Sampling a distribution of strongly correlated variables is a central
task in many fields such as statistical mechanics, machine learning
and statistical analysis. Indeed, there are many problems where an exact treatment
is impossible due to the large number of strongly correlated variables.
Furthermore, in many cases analytical approximations lead to imprecise results if compared to the one obtain in numerical simulation. The Markov Chain Monte-Carlo (MCMC) approach
for sampling is a fundamental component of modern physics
\cite{binder1986monte,krauth2006statistical} playing a central role
in inference and learning problems (e.g.  computational
biology \cite{BIOM:BIOM1,o2002tutorial}, machine learning
\cite{gamerman2006markov}, simulated annealing
\cite{Kirkpatrick13051983}, ...). A drawback of MCMC methods, such as
Metropolis, is the long runtime needed to obtain high-precision
estimates. In addition, this time can be affected by local energy or
entropy barriers and ergodicity breaking. A large scientific effort
has been devoted to the design of faster MCMC
schemes \cite{binder1986monte,krauth2006statistical}.

A particular family of discrete statistical models considers systems where
the underlying graph of interaction is a tree. Those cases have been
widely studied both in physics, where they form the basis of the Bethe
approximation \cite{MezardParisi87b}, and in computer science
\cite{mackay2003information,Pearl82}. In these problems, an {\it exact}
marginalization in linear time is possible --- i.e. in
$\mathcal{O}(N)$ steps, where $N$ is the system size--- by using an
algorithm called Belief Propagation (BP). As we shall see, it implies
the possibility of perfect sampling in linear time as well (i.e extracting a configuration
from the Boltzmann measure). This has been a
fundamental breakthrough allowing gigantic possibilities in machine
learning \cite{pearl1988probabilistic}. A natural extension is to
consider graphs % sampling the measure for graphs
 with few (or large) loops. Those graphs are typically present in many concrete problems %AD check the sentence
\cite{wainwright2008graphical}. But when the loops appear it becomes
%as soon as loops appear it becomes
much harder to sample perfectly the measure. It would be indeed very useful to
be able to sample efficiently graphs that have {\it locally} a tree
structure but which are {\it not} tree. Among such graphs, random ones
are commonly used in statistical physics as mean field
models \cite{MezardParisi01}, in combinatorial optimization as
archetypes of hard benchmarks \cite{MonassonZecchina99}; and in many
types of sparse networks	 encountered in clustering
problems \cite{decelle2011inference,girvan2002community}. They are also
used for error correcting codes \cite{richardson2001capacity} and in
inference problems \cite{wainwright2008graphical}, where a good sampler
is mandatory if one hopes to deal with large size problems.

\paragraph*{Belief Propagation Guided MCMC --- } 
In this paper we present an algorithm respecting the detailed-balance which combines the perfect sampling ability of BP on trees with the traditional heat-bath strategy using MCMC approaches. A method close to our has been studied on 2D lattices in \cite{matsubara1997cluster}. However our algorithm is able, on random graphs, to flip random trees of huge sizes which would have been impossible with their method.  Those very large clusters allow to avoid local traps that one can encounter with the local Metropolis dynamics. We expect in addition that our algorithm unleashes its
full potential on graphs without too many short loops where the sub-tree
extraction is facilitated. Finally, our algorithm can be adapted for any kind of graph.
A similar approach has been used in \cite{hamze2004fields} but where only deterministic spanning trees were considered.
%It is expected to fully unleash its potential on graphs without too many short loops. 

In what follows, we present the algorithm
in detail and apply it to various difficult benchmarks. We focus mainly
on systems having the random graph topology, as they are typical
examples of networks without short loops, and we demonstrate the
state-of-the-art nature of our approach.

\begin{figure*}[t]
  \includegraphics[width=4.8cm]{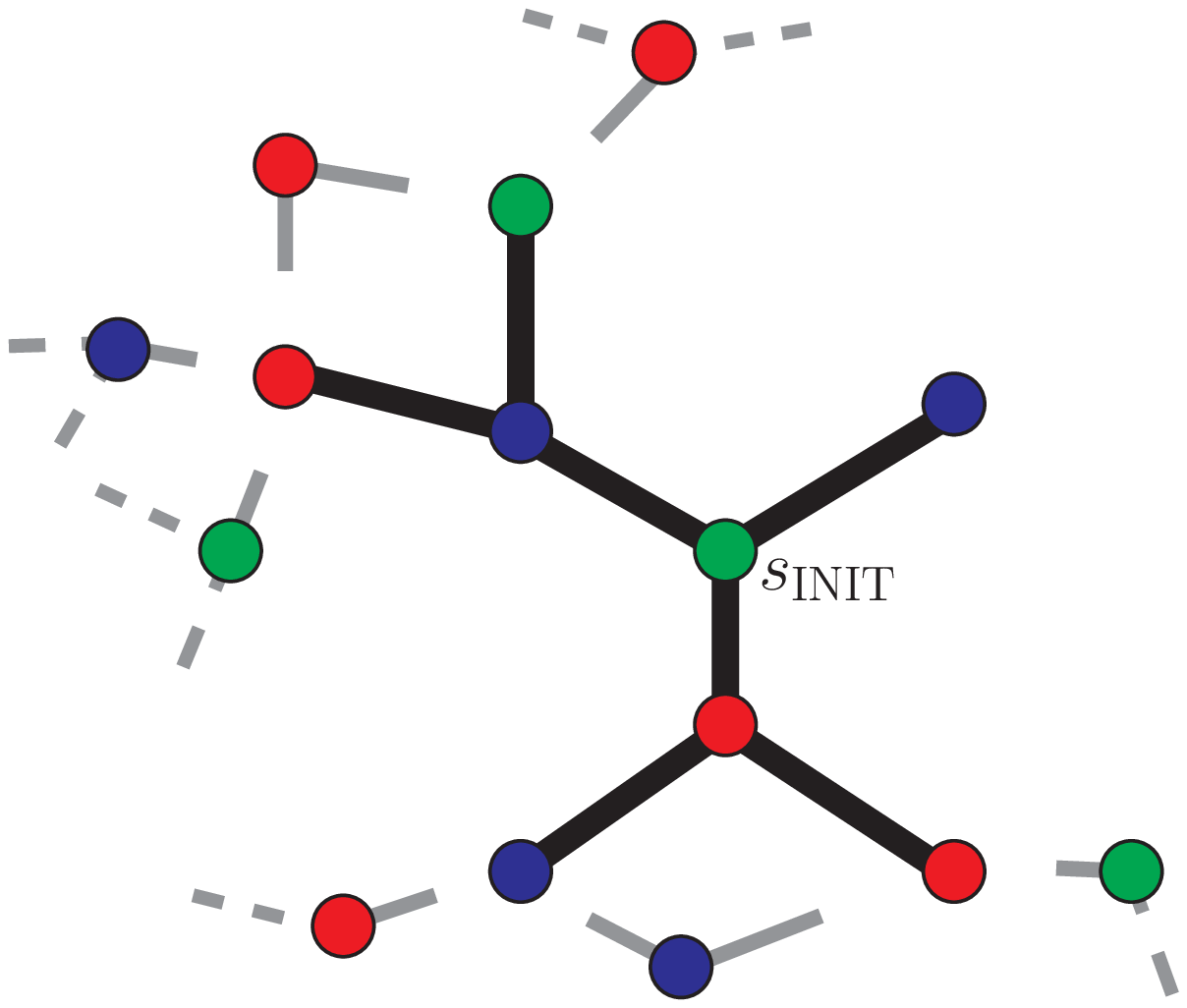}\hspace{1cm}
  \includegraphics[width=5.3cm]{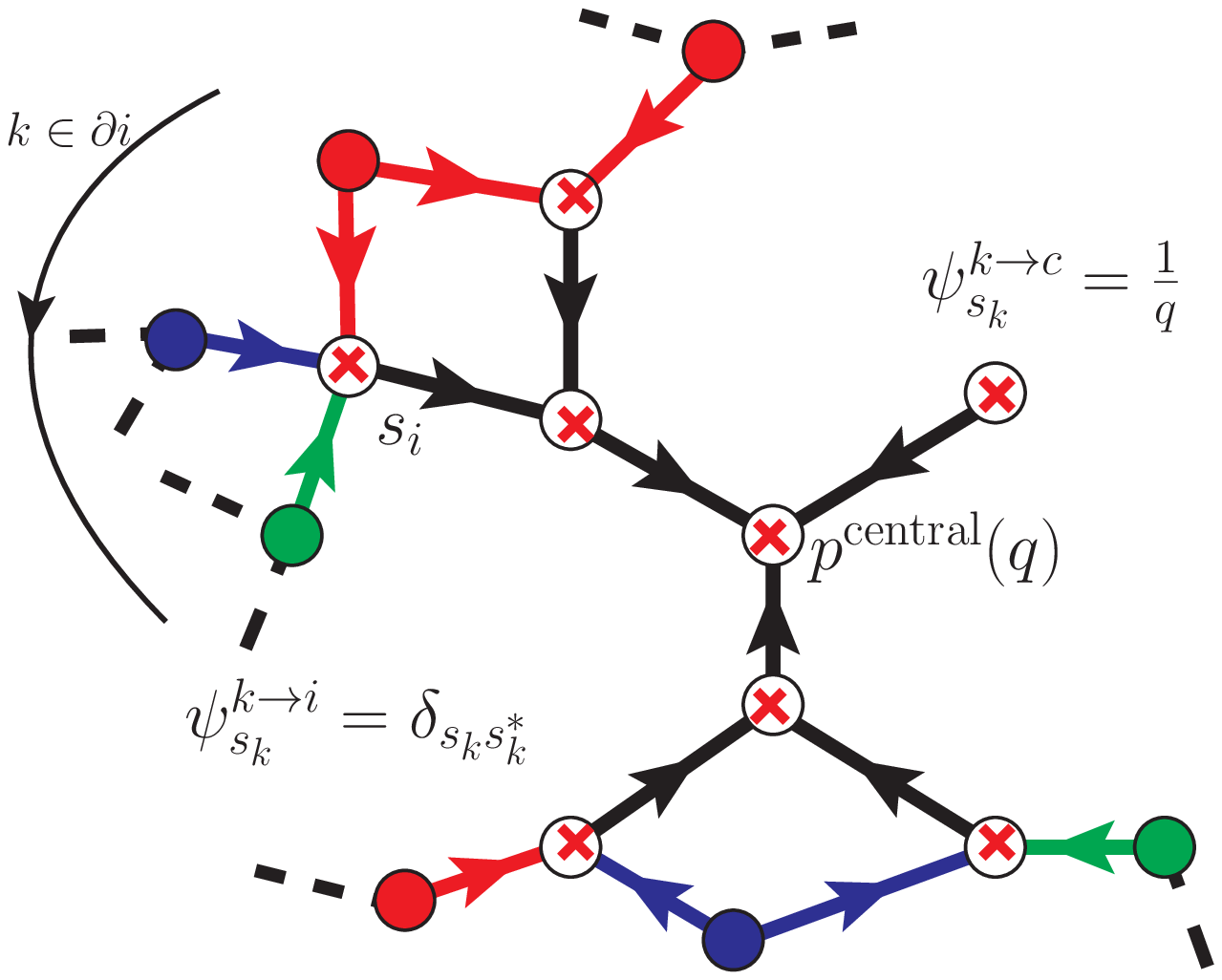}\hspace{1cm}
  \includegraphics[width=5.0cm]{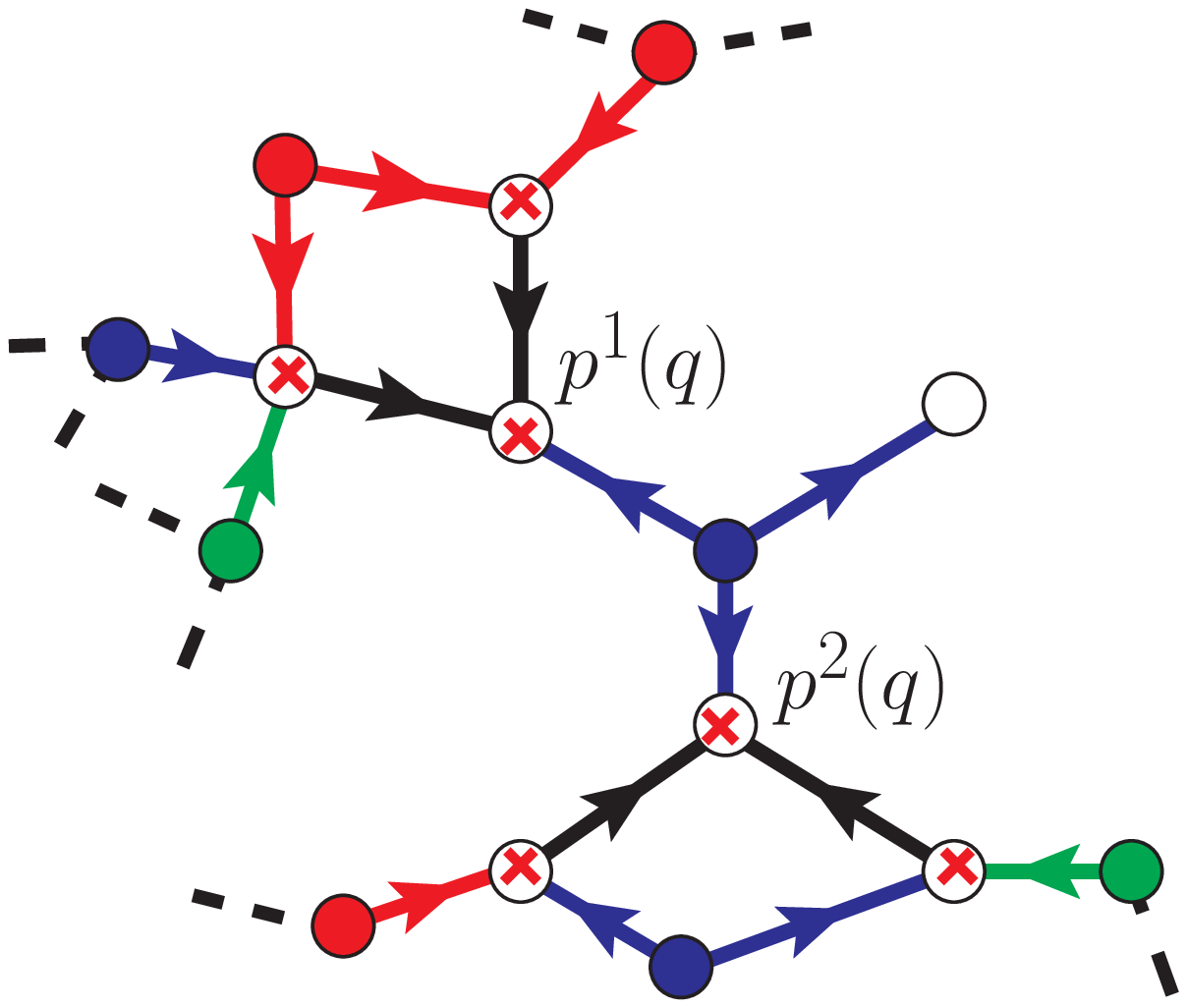}
  \caption{A schematic illustration of the BP-guided
    MCMC on the coloring problem. \textbf{Left:} Starting from a random node $s_{\rm INIT}$ we create a random tree by 
    adding the neighbors of each node in the tree progressively (but without creating a loop).
    The (plain)
    links in grey have been cut on the picture to emphasis the constructed
    tree-structure.  \textbf{Middle:} the tree nodes have been resetted
    and marked by a red X on the figure. We illustrate by arrows how the BP messages
    propagate to compute the partial marginals until the central spin is reached. For the latter we can then compute the complete marginal. \textbf{Right:} the
    central node has been put in one state according its marginal
    $p^{\rm central}$. Propagating this information backward on the
    tree, this allows to compute the complete marginal for each
    variable and to sample a new configuration for all the variables
    on the tree.} \label{fig_ext1}
\end{figure*}

Our method is based 
on a heat-bath procedure: we repeatedly select a random sub-part of the
problem and equilibrate it given its interaction with the rest of the
system. When it is applied to a single spin, it leads to
Glauber dynamics. A common strategy to improve the convergence of the dynamics is to apply it to a group of two or more spins. However, the difficulty to perform a perfect
sampling increases dramatically with the number of spins in general. Our
strategy will be to select sub-parts of the network having the topology of a tree and to
use BP to guide our sampling process. 

\paragraph*{Sampling on a tree ---} We describe first how using BP
we can sample efficiently the Boltzmann measure on a tree given the
boundary conditions. Starting from the leaves of the graph (the nodes with only
one neighbor), we can
compute sequentially the BP messages (or ``partial'' marginals) toward the center. Let's consider first
a ``leaf'' spin $s_l$ and compute its ``partial'' marginal toward $k$, that is, 
the probability of $s_l$ given that its only neighbor $k$ has been removed.	
When the only neighbor has been removed, the ``leaf'' spin is only sensitive 
to the boundary condition (which we
denote as an effective magnetic field $h_{i}^{\rm bound}$) and the
partial marginal reads
\begin{equation}
  \psi^{l \rightarrow k}_{s_l} = \frac{e^{\beta h^{\rm bound}_{l} (s_{l})}}{Z^{l \rightarrow k}} 
\end{equation}
\noindent where $Z^{l \rightarrow k}$ is the normalization constant 
and $\beta$ the inverse temperature. In the
following, we should also consider the partial marginal $\psi^{i \rightarrow j}_{s_i}$ of a site $i$ when the link $(ij)$ has been removed (also called BP message from $i$ to $j$).
Starting from the leaves, these partial marginals can be propagated toward the center
of the tree using the BP equations: for any $i$, $\psi^{i \rightarrow j}$ can be computed when the $\psi^{k \rightarrow i}$ are known for all neighbors $k\neq j$ of $i$. This propagation is exact on trees and reads for $\psi^{i \rightarrow j}$:
\be
	\psi_{s_i}^{i \rightarrow j}=\frac{1}{Z^{i \rightarrow j}} \prod_{k \in \partial i \backslash j} \sum_{s_k} e^{-\beta \mathcal{H}(s_i,s_k)} \psi_{s_k}^{k \rightarrow i}
\ee
\noindent where $\mathcal{H}$ encodes the interaction between
neighbors, and $\partial i$ denote the neighbors of $i$. 
Iterating this procedure allows to go deeper and deeper in
the tree, until we reach a spin where {\it all} incoming messages have been computed.
 For this very last spin, we can now
compute the correct ``complete'' marginal using
\be
  p_{\rm marg}(s_c) = \frac{1}{Z^c}\prod_{k \in \partial c} \sum_{s_k} \psi_{s_k}^{k \rightarrow  c} e^{-\beta \mathcal{H}(s_k,s_c)}
\ee
Thus, we can use this marginal to choose a new state for the spin $s_c$.
% choose a new state for this spin accordingly to its marginal.
Given this new assigned value for the spin's state, it is possible to
compute the complete marginal for its neighbors. From them, we choose
again a new state for these spins, and this procedure is iterated back to
the boundaries of the tree. At this point, the whole tree has been
updated with a new configuration sampled from the Boltzmann measure in $\mathcal{O}(N)$ steps, where $N$ is the size of the tree graph.

\paragraph*{Sampling on a graph ---} We now explain how to use this procedure to
perform a heat bath on any graph. % ( as illustrated on
%FIG.~{\ref{fig_ext1}. 
  The procedure follows three different steps: (i)
  Extract randomly a tree sub-graph from the
  interaction graph; %AD of the model considered; 
  see FIG.~\ref{fig_ext1} left
  panel. (ii) Cancel the states of the spins inside the tree. All spins
  {\it immediately outside} of the tree will be used as the boundary
  conditions. (iii) Use the BP perfect sampler described above to
  extract a new configuration of the spins {\it inside} the tree given
  the boundary conditions. On FIG.~\ref{fig_ext1} the middle panel
  illustrates the propagation of BP messages toward the central spin.
  On the right panel, how new states are drawn and used to iterate
  BP messages.

  In our implementation of the algorithm, we construct the sub-tree by
  taking a node at random and adding its neighbors in random order. The neighbors are added 
   unless it creates a loop in the sub-graph. In such case we put 
  it in the list of spins at the border of the tree. This list will be used as boundary conditions.
  The procedure is repeated on all newly added nodes until all of them have been treated. 
%  its neighbors iteratively until
%  it creates a loop. Denoting SL the sub-tree and NL its frontier, the
%  following
%  pseudo-code describes this process:
%
%%
%\begin{codebox}
%\Procname{$\proc{Tree-Maker}(G)$}
%\li $SL=\emptyset$ and $NL=\emptyset$
%\li Take a site $s_{\rm INIT}$ at random in the graph $G$
%\li $SL = SL \cup s_{\rm INIT}$
%\li \For ($s \in SL$)
%\li		\Do : \For( $s_n \in \partial s$ )
%\li		 	\If ($\forall i \in \partial s_n \backslash s$, $i \notin SL$)
%\li 			\Then $SL=SL \cup s_n$ 
%\li			\Else $NL=NL \cup s_n$		
%	\End \End \End
%\li \Return SL and NL
%\end{codebox}
%
%\noindent The TreeMaker method is expected to be very efficient in
%graph with few short loops and therefore very effective on random
%graphs. 
This construction is particularly efficient on random graphs and we
illustrate it on the left panel of FIG.~\ref{fig_ext1}. It is however important
to point out that it might not be efficient on finite dimensional systems. 
In that case one should 	design a different procedure to extract
 a tree from the graph.

%However, one should
%be careful that on finite dimensional systems it might be needed to
%design a different procedure to extract a tree from the graph.
%In the case of finite dimensional systems it might be needed to
%design a different procedure to extract a tree from the graph. An
%example of this procedure is illustrated on the left panel of
%FIG.~\ref{fig_ext1}. %In order to not construct always the same graph
%when starting from one node we added randomness in the sub-tree construction by
%selecting neighbors in random order or by adding a node with a given
%probability. 
The creation of the sub-tree is dominating the
algorithm's complexity and a complete update of the graph scales as
$\mathcal{O}(c^2 N)$, where $c$ is the average degree of the graph.
Our algorithm is therefore faster on diluted graphs. %Its complexity
%is higher by only a factor $c$ compare to a Metropolis algorithm. 
However, we should be careful that in some cases this construction might update less frequently some sites. 
Indeed, since the root of the tree is chosen randomly, small disconnected cluster will be updated more rarely. To counterbalance  this effect, we always alternate our algorithm with a Metropolis move so that all the sites are updated frequently. It is also important 
to point out that this is not a problem of ergodicity: all states of the
phase space can be reached with a non zero probability.
In our tests of the algorithm, the results were mainly independent of the amount of
randomness used during the sub-tree creation. We also observed that
the local MCMC moves were quite important when dealing with Poissonian
random graphs.

\paragraph*{Numerical tests --- } We shall now discuss the performance of our
algorithm. In the following we consider three different examples. First we focus our attention
on the relaxation of the energy as a function of the time. We compare three different methods
on two Ising models and we investigate the effect of the sub-tree maximum size on the convergence. Second we study the auto-correlation time for both metropolis and our method on a $p$-spin. Finally we compare the same algorithms on an annealing experiment. Note that for our algorithm one time-step corresponds to an update of $N$ spins ($N$ being the system's size) such that after $T$ time-steps all sites are updated $T$ times in average. This definition is chosen in order to make a fair comparison with MC where a time-step corresponds to the update of $N$ randomly chosen spins. We should also mention that the results presented here do not depend on the system sizes. We controlled that using larger (or smaller) system sizes the conclusions were not affected. 

We consider first the energy relaxation after a
quench in the low temperature phase of two systems. We consider an easy case ---a
standard ferromagnetic Ising model on a random graph with connectivity
$c=3$, $T_c\approx 2.94$ --- and a harder one --- an anti-ferromagnet on the same random graph, $T_c\approx 1.52$. This later, 
due to the presence of loops of various sizes, behaves as a spin
glass model. For both, we start from a random 
configuration and cool the system at $T=0.1$. A quench from a random configuration at this temperature get stuck into relatively high energy states due to the presence of local energy barriers. 

\begin{figure}[t] \includegraphics[width=8.5cm]{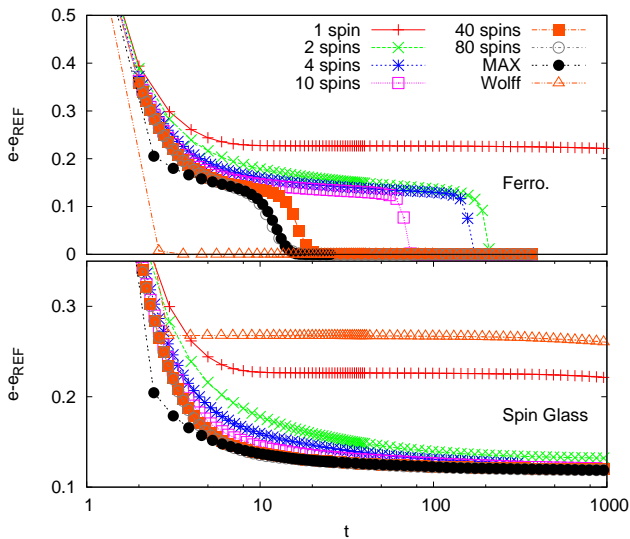}
  \caption{Energy density after a quench (using $e_{REF}=c/2$ for the ferromagnet and $e_{REF}=0$ for the antiferromagnet) starting
    from a random initial configuration for an Ising model on a large
    ($N=10^6$) regular random graph with connectivity $c=3$ at
    temperature $T=0.1$. \textbf{Top panel:} The convergence of the
    energy is shown versus the iteration time in the ferromagnetic
    case using different thresholds for the largest possible
    tree-cluster move on each curve. While standard metropolis ($1$
    spin) get trapped in configurations with large energy for
    infinitely long time, increasing the size of the trees
    systematically increases the efficiency of the algorithm. With
    large enough trees, one equilibrates the system in less than $20$
    iterations. \textbf{Bottom panel}: A similar study for an
    antiferromagnet that behaves as a spin glass model due to the
    frustration. Similar performances as in the ferromagnetic case are
    observed. Note that while the Wolff cluster approach is able to
    perfectly sample the ferromagnet, it fails in the spin glass case
    where the best performances are obtained by the BP-guided
    algorithm.} \label{fig_ferro} \end{figure}

In our simulation, we compare the relaxation time of our algorithm where we add a parameter controlling the sub-trees's maximum size and we report the results obtained by varying this parameter (when the maximum size is one we recover the Glauber MCMC). In addition we implemented the Wolff algorithm \cite{wolff1989comparison} to confront a cluster method to the BP-guided one. On the FIG.~\ref{fig_ferro} we plot the energy as a function of
the iteration time for both systems. As we increase the maximum size of the
sub-trees, we update larger and larger clusters and the
barriers no longer block the dynamics. One can observe on the figure that 
the convergence time is drastically improved.
In fact, perhaps not surprisingly, the algorithm
converges faster when using the maximum cluster size ---thus avoiding
larger and larger local minimas--- in both the ferromagnet and the
spin glass case. It is instructive to compare it with the standard Wolff algorithm. For the ferromagnet,
Wolff is able to avoid the barriers and in fact performs
even better than the BP-guided MCMC. This is hardly a surprise as for
ferromagnetic systems without frustration the Wolff approach is always very efficient. 
For the spin glass problem, however, even the Wolff method
remains stuck in some sub-space of the phase space (see
FIG.~\ref{fig_ferro}). In these cases, the advantage of our approach is
evident.

%
%When allowing only one spin, we recover the standard Glauber MCMC. At this temperaturen both systems, at this
%temperature, a quench from a random configuration get stuck into
%relatively high energies due to the presence of local energy
%barriers. We plot on FIG.~\ref{fig_ferro} the energy as a function of
%the iteration time for both systems. As we increase the size of the
%updated tree, we start to update larger and larger clusters and the
%barriers no longer block the dynamics. This improves the convergence
%time drastically. In fact, perhaps no surprisingly, the algorithm
%converges faster when using the maximum clusters size ---thus avoiding
%larger and larger local minimas--- in both the ferromagnet and the
%spin glass case. It is instructive to compare with the standard Wolff
%Cluster \cite{wolff1989comparison} algorithm. For the ferromagnet,
%Wolff algoritmh is able to avoid the barriers and in fact performs
%even better than the BP-guided MCMC. This is hardly a surprise, given
%how good the Wolff approach is for ferromagnetic systems without
%frustration. For the spin glass problem, however, even the Wolff
%approach remains stuck in some sub-space of the phase space (See
%FIG.~\ref{fig_ferro}). There, the advantages of our approach is
%evident.

\begin{figure}[t] \includegraphics[width=8cm]{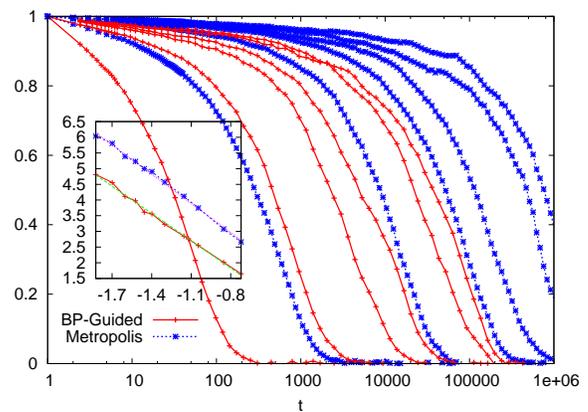} % AD check the legend
  \caption{Autocorrelation $m(t)=C_{\rm eq}(t)$ for the $p$-spin on
    regular graph with $N=10^5$ and $c=3$ starting from
    equilibrium. The temperatures go from $0.7 \rightarrow 0.525$ and
    the dynamic glass transition arises at $T_d=0.510$
    \cite{MontanariSemerjian06b}. In the inset the melting time
    diverges as a power law. The BP-guided MCMC method is more than $10$
    times faster than the traditional one.}
\label{fig_fast_quench}
\end{figure}

As a second example, we move to a more complicated model: the Ising
$p$-spin glass on random graphs, also known as the random XORSAT model. The
Hamiltonian reads
\be
	\mathcal{H}(\{s\})= -\sum_{a=1}^M J_a \prod_{i \in \partial a} s_i,
\ee
where $M$ groups of $p$ spins are chosen randomly and coupled according
to the random coupling $J_a$ which is equal to $\{\pm 1\}$ with equal
probability.
This model has been widely studied in the literature both as model for
glasses \cite{derrida1980random,franz2007ferromagnet}, for error
correcting codes \cite{mezard:10} and as a toy model for the
satisfiability problem \cite{ricci2001simplest}. It exhibits a
dynamical glass transition at $T_d = 0.510$
\cite{MontanariSemerjian06b} at which the relaxation time diverges, thus making difficult to 
take independent measures. We study here the relaxation 
time of the magnetization when $T \rightarrow T_d$ for both Metropolis and our algorithm. By using the approach of
\cite{MontanariSemerjian06,krzakala2011melting}, it is possible to
create perfectly thermalized initial conditions which allow us to
measure the equilibrium spin-glass correlation time easily. The time-relaxation
of the correlation as $T \rightarrow T_d$ can be directly observed starting 
from this thermalized configuration. In FIG.~\ref{fig_fast_quench} we illustrate the alpha relaxation
starting from an equilibrium configuration for different temperatures
close to $T_d$. Our algorithm improves the
decorrelation time by one-order of magnitude while the exponent for
the diverging relaxation time was, however, not significantly
smaller. Hence, even in models as complicated as the XORSAT one, the BP-guided approach improves the mixing property of the dynamics by taking advantage of the local-tree structure.

A third example is given by the q-coloring problem. This is a NP-complete
constraint optimization problem that aims to color a graph with $q$
colors such that all variables have a different color than their
neighbors. It is equivalent to an antiferromagnetic Potts model:
\be \mathcal{H}(\{s\})=  \sum_{(i,j)
  \in \mathcal{E}} \delta(s_i,s_j) \, , \text{~with~$s_i=1,...,q$.}\ee

\begin{figure}[t] \includegraphics[width=8cm]{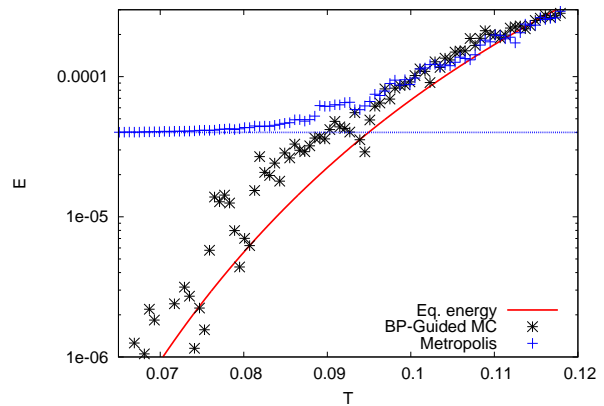}
  \caption{Simulated annealing starting from an equilibrated
    configuration at $T=0.12$ in a 4-coloring problem on a large
    $N=10^5$ 9-regular random graph. The annealing is performed by decreasing the temperature by
    $\Delta T=10^{-7}$ every ten time-steps.  While the traditional
    metropolis approach is stuck at finite energy, our BP-guided
    algorithm shows no sign of such blocking and actually reaches the
    ground state.}\label{fig_anneal_canyon}
\end{figure}

\noindent We consider this model on random graphs for which, in some region of the parameter $q$ and the average degree $c$, a coloring configuration exists with probability one but can be hard to find. In a recent work \cite{ZdeborovaKrzakala10}, it has been shown that the $9$-regular graph with $q=4$ has a dynamical transition below which the equilibrium states possess a colorable configuration. We therefore perform an annealing experiment from an equilibrium configuration at $T<T_d$ using MCMC dynamics on one part and the BP-guided approach on the other one. As can be seen in FIG.~\ref{fig_anneal_canyon}, the MCMC dynamics gets stuck in some local minima as the temperature is cooled down. However, under the same condition, our algorithm manages to escape of such minima and to reach the ground state of the system.

%\noindent We perform a thermal annealing ---probably the most used
%technics in optimization \cite{Kirkpatrick13051983}--- in order to
%find the ground state. To show how hard the problem is, we performed 
%an annealing in a $9$-regular graph with $q=4$ colors which
% have a dynamical transition at $T=0.15<T_d$\cite{KrzakalaZdeborova07}. 
% Even if one is starting from
%an equilibrium configuration at a temperature $T=0.12<T_d$ a local
%MCMC dynamics gets stuck in some local minima as the temperature is
%cooled down. However, in the same condition (see
%FIG.~\ref{fig_anneal_canyon}), the BP-guided approach manages to
%escape such minima and to reach the ground state.
%

\paragraph*{Conclusion --- } We have presented a new algorithm for
exact sampling in complex systems, illustrated its performances and
compared it to those of more traditional Metropolis dynamics. We
show different examples where our method performs better than local
move MCMC. In addition we demonstrate that our algorithm out-competes
(some) cluster rejection-free method and is immediately adaptable to
many types of systems. We have made all the tests on random graphs
since the extraction method of sub-trees we are using is particularly
adapted for this topology. On the other hand, the algorithm can be
applied to any kind of graph (except fully-connected ones), but one
should be careful when choosing the sub-graph construction. Indeed for
a typical Euclidean graph the first step of the algorithm has to be
optimized in order to construct quickly a sub-tree for the network
considered. Many developments could be considered: combining our
algorithm together with parallel tempering; testing the performances
on finite dimensional models, as for instance the diluted spin models
of \cite{PhysRevLett.100.197202} where large trees could be
constructed; or studying the zero temperature version as an
optimization tool.

\paragraph{Acknowledgments ---}
This work has been supported in part by the ERC under the European
Union's 7th Framework Programme Grant Agreement 307087-SPARCS. 
A. Decelle has been supported by the FIRB project n. RBFR086NN1.

%--------------------------------------------------------------------------

\bibliography{myentries}	

\end{document}